\documentclass[12pt,letterpaper]{article}
\usepackage[a4paper, total={7in, 10in}]{geometry}

\usepackage{graphicx}
\usepackage{helvet}
\usepackage{authblk}
\usepackage{hyperref}
\usepackage{amsmath} 
\usepackage{amssymb} 
\usepackage{mathptmx}
\usepackage{mathrsfs}
\usepackage{orcidlink} 
\usepackage[super,comma,sort&compress]  
   {natbib}
\usepackage[right]{lineno} \linenumbers

\makeatletter
\renewcommand{\maketitle}{\bgroup\setlength{\parindent}{0pt}
\begin{flushleft}
  \textbf{\@title}
  
  \@author
\end{flushleft}\egroup}
\makeatother

\title{Lattice physics approaches for neural networks}
\date{}

\author[1,3,*,\orcidlink{0000-0002-3731-8202}]{Giampiero Bardella}
\author[1**,\orcidlink{0000-0003-3304-3157}]{Simone Franchini}
\author[1,2,\orcidlink{0000-0001-6825-1162}]{Pierpaolo Pani}
\author[1,2,\orcidlink{0000-0001-6100-2438}]{Stefano Ferraina}

\affil[1]{Department of Physiology and Pharmacology, Sapienza University of Rome, Italy}
\affil[2]{Equally contributing co-senior authors}
\affil[*]{Correspondence: giampiero.bardella@uniroma1.it}
\affil[**]{Correspondence: simone.franchini@yahoo.it}
\affil[3]{Lead contact}

\begin{document}

\maketitle

\section*{Abstract}

Modern neuroscience has evolved into a frontier field that draws on numerous disciplines, resulting in the flourishing of novel conceptual frames primarily inspired by physics and complex systems science. Contributing in this direction, we recently introduced a mathematical framework to describe the spatiotemporal interactions of systems of neurons using lattice field theory, the reference paradigm for theoretical particle physics. In this note, we provide a concise summary of the basics of the theory, aiming to be intuitive to the interdisciplinary neuroscience community. We contextualize our methods, illustrating how to readily connect the parameters of our formulation to experimental variables using well-known renormalization procedures. This synopsis yields the key concepts needed to describe neural networks using lattice physics. Such classes of methods are attention-worthy in an era of blistering improvements in numerical computations, as they can facilitate relating the observation of neural activity to generative models underpinned by physical principles.\\

\noindent\textbf{Keywords:} neural networks; generative models; statistical physics; lattice field theory; entropy; neurophysiology; network inference; bayesian inference; free energy principle; active inference;


\section*{Introduction}

Systems neuroscience aims to understand how the spatiotemporal interaction of aggregates of neurons leads to neural dynamics and, ultimately, to cognitive processes and behavior. The ever-increasing technological innovations in the design of high-resolution recording devices have given a dramatic boost to data gathering, and we can now simultaneously sample the activity of hundreds of neurons with outstanding temporal resolution\cite{Angotzi2019,Musk2019,Obaid2020,Steinmetz2021,Normann1999,Leber2017,Ye2024}. Yet, the immense number of observations still fails to provide conclusive evidence on the mechanisms that govern neural systems, the link between connectivity and dynamics, and the emergence of composite behavioral functions. To progress, neuroscience needs to start learning governing principles from data. To conquer this obstacle, a systematic and methodical integration of experiments, theory, and computational modeling is indispensable. Neuroscience is now where particle physics was before the introduction of the Standard Model. It is still common to heavily rely on heuristic analysis and modeling approaches that only partially account for the richness of the spatiotemporal repertoire of neural states. 
Despite the proposal of several landmark models\cite{Beurle1956,Amari1972,Wilson1972,Fischer1973,Amari1977,Hopfield1982,Amit1985,Amit1997,Toulouse1986,Treves1991,Abeles1982,Abeles1995,Buice2007,Buice2013}, we are still far from achieving a mechanistic understanding of the nervous system through deep physical principles\cite{Wigner1961,Weizcsacker1984,Penrose1999}. In a recent paper\cite{BardellaLFT2024}, we introduced methods from theoretical particle physics and quantum field theory (QFT\cite{Brown2005,LSZ1955,Guerra1975,Parisi1981, Damgaard1987, Parisi1989,Lee1983,Lee1987,Rovelli1995,Hooft2014,Wilson1974,Wiese2009, Gupta2011, Zohar2015, Parotto2018,Magnifico2021,Gornitz1992,Peretto2004,Deutsch2004,Singh2020,Franchini2023, DiCastro1969,Balian1974,Wilson1983,Parisi2001,Kadanoff2013,Franchini2021})
to treat systems of interacting binary variables, like the spiking activity of interacting neurons in the brain. The intuitive mathematical formalism\cite{Franchini2023} allows the explicit interpretation of neural interactions through universal laws, reducing the gap between abstraction and experiments thanks to the direct connection between model parameters and experimental observables. 
Such a feature may facilitate our understanding of systems characterized by computational complexity, such as the brain at a fine-grained neuronal level, which requires approaches that allow for approximate solutions. Some of these approaches exploit a probabilistic perspective on brain computations based on methods such as Bayesian inference\cite{Pouget2013,Friston2010,Gentili2021} and fuzzy logic\cite{Zadeh1964}. Importantly, the brain's fuzzy logical characteristic has already been suggested at the functional anatomical level, particularly in sensory systems (i.e., cortical columnar organization)\cite{Gentili2021}. Indeed, cortical activity and neuronal features, such as receptive fields, have been shown to exhibit overlapping and gradual boundaries, thus displaying fuzzy properties\cite{Gentili2021}.
These fuzzy properties are also evident at the abstract level of cognitive processes, i.e., when considering the ability of the brain to “compute with words” in uncertain context.\cite{Zadeh1999} Together, Bayesian and fuzzy logic aim to explain the high-level key features of the brain, including its operation through predictions in encoding incoming sensory information, its ability to elaborate motor plans for interacting with the environment, and its capacity to make decisions in uncertain situations. Importantly, in appropriate experimental settings, we can measure these predictions and decisions solely based on overt behavior. Using our physics-guided formalism on neural data would enable a new class of generative models of neural dynamics, providing additional basis for benchmarking some unifying brain theories\cite{Pouget2013,Gentili2021} relying on a Bayesian perspective, such as the free energy principle of Friston et al.\cite{Friston2010,Fagerholm2021,Zeidmann2023}.

\section*{Background}

In computational neuroscience, two visions are currently dominant: manifold and circuit modeling. The first postulates that embedding the high-dimensional state space of neural dynamics into low-dimensional surfaces, i.e., manifolds, reveals neural computations, whereas the latter considers connectivity among neural units as a founding mechanism. In broad terms, we can surmise that neural manifolds provide a descriptive modeling of neural activity based on statistical interdependencies without links to causal mechanisms. That is because to define low-dimensional subspaces in empirical data, a set of numerical dimensionality reduction steps is required, which are most of the time arbitrary and not unique.  An example is principal component analysis (PCA), which seeks low-dimensional projections that maximize experimental data variance. These methods have proven to be efficient not only for exploratory analysis and as denoising tools but also for providing satisfactory predictions of experimental variables in certain cases\cite{Kobak2016,Murray2017,Gallego2017,Jazayeri2021,Vyas2020,Pani2022}.
However, they systematically yield contradictory conclusions that are not interpretable under comprehensive principles and several criticisms have recently arisen about the unreliability of these methods' interpretive power\cite{Lebedev2019,Shinn2023,Kuzmina2024}.
An example is the formal relationship that these methods can establish with physical quantities. The most suggestive one for neuroscientists is the correspondence between the set of possible configurations of a neural system and the physical notion of energy, which remains rickety, as in the case of the PCA-based "energy landscape"\cite{Gallego2017,Langdon2023,Wang2023}. On the other hand, circuit models, despite their ability to integrate causal assumptions through a wide range of realistic biophysical parameters (membrane potential, cell types, etc.), typically rely on specific tuning of model parameters to replicate observations. Although this class of model showed success in predicting experimental variables\cite{Mante2013,Wimmer2014,Sussillo2015,Rajan2016,Finkelstein2021},performing specific computations\cite{Mastrogiuseppe2018} and replicating neural dynamics evolving on low-dimensional manifolds\cite{Langdon2023,Pollock2020}, the frequent shortcomings are flattened stimulus selectivity\cite{WangXiao2002}, uniform time evolution of reconstructed activity\cite{Langdon2023} and tremendous difficulties in scaling the analytical treatments for networks of arbitrary sizes\cite{Fasoli2016,Brinkman2022}. Moreover, even when resolving the problem of homogeneous stimulus tuning, accounting for activity modulated by multiple independent variables, the so-called mixed selectivity\cite{Murray2017,Chaudhuri2019}, the crafted connectivity of such models has a blurred link with predicted activity and real circuit mechanisms. Extensive discussions on the numerous open problems of how connectivity leads to time-dependent activity can be found in the works of Barack and Krakauer\cite{Barack2021} and Langdon\cite{Langdon2023} and colleagues. Although the manifold and circuit modelling strategies are the most popular, their fragmented view falls short of formally reconciling with general physical principles\cite{Friston2010}. One of the most exhaustive ways of describing brain spatiotemporal dynamics is neural field theory\cite{Buice2007,Deco2008,Buice2013,Qiu2014,Nanopoulos1995,Friston2010,Crisanti2018,Helias2020, Gosselin2020,Fagerholm2021,Halverson2021,Summers2021,Swan2022,Tiberi2022,Cook2022, Peretto1984,Schneidman2006,Meshulam2024,BardellaLFT2024}. Usually, it models large-scale average population activity in the continuum limit, including approximated anatomical and physiological details into differential equations, with the Wilson-Cowan model\cite{Wilson1973} being probably the most popular example. As thoroughly discussed in a recent review on the topic\cite{Cook2022}, many types of approaches fall under this definition, some of which may overlap with those discussed above. However, in addition to the high number of parameters that are not always attributable to experimental variables, their main limitation is the lack of a consistent mathematical framework to link the various scales at which neural dynamics unfolds. One of the most successful ways to embed physical laws into neuroscience modeling has been by adapting methods from equilibrium statistical physics to real and artificial neural networks. Since the introduction of the Amari-Hopfield model\cite{Amari1977,Hopfield1982,Amit1985,Amit1997,Toulouse1986,Treves1988,Treves1991,Abeles1995,Schneidman2006}, the most fruitful formalism comes from the physics of magnetic systems. In this language, the energy function parallelism is realized through the analogy with the energy of a spin system. 
As accurately described by Brinkman\cite{Brinkman2022} and colleagues, extensions to time-dependent non-equilibrium dynamical systems have mostly involved the use of stochastic dynamics (in two equivalent forms, the Fokker-Planck equation\cite{Brunel1999,Fusi1999,Vinci2023,Vinci2024} and the Langevin equation\cite{Yan2013,Wang2014,Yan2020,Shi2022,Genkin2021,Shi2023}), stochastic variations of the Ising model\cite{Tyrcha2013,Aguilera2021} and semi-analytical studies based on bifurcation theory\cite{Haschke2006,Fasoli2019}. The major weakness of these formulations is that exact solutions could become hard to obtain, as could a precise and simple mapping between theory, simulations and empirical data. 
Several authors\cite{Wigner1961,Weizcsacker1984,Penrose1999,Buice2007, Buice2013,Qiu2014,Nanopoulos1995,Friston2010,Crisanti2018,Helias2020,Gosselin2020,Fagerholm2021,Halverson2021,Summers2021,Swan2022,Tiberi2022,Cook2022,Peretto1984,Schneidman2006,Meshulam2024,BardellaLFT2024,Dick2024}have already argued that the most promising strategy to overcome all of these shortcomings is to adopt a different perspective, pursuing the formalism of QFT\cite{Brown2005,LSZ1955,Guerra1975,Parisi1981,Damgaard1987,Parisi1989,Lee1983, Lee1987,Rovelli1995,Hooft2014,Wilson1974,Wiese2009,Gupta2011,Zohar2015,Parotto2018,Magnifico2021,Gornitz1992,Peretto2004,Deutsch2004,Singh2020,Franchini2023,DiCastro1969, Balian1974,Wilson1983,Parisi2001,Kadanoff2013,Franchini2021}. 

\section*{Preliminaries}
Let us indulge in a few informal remarks on the Lagrangian description of a dynamical system, pivotal to our formulation. Except among physicists, the Lagrangian interpretation is less known compared to other approaches, such as Newtonian mechanics, even if it is more foundational because it considers only energy, generalized coordinates, symmetries and conservation laws. A Lagrangian treatment is particularly useful for complex systems with multiple degrees of freedom, providing a unified framework connecting almost every phenomenon in nature, from classical mechanics to electromagnetism and field equations in general relativity.  It assumes the existence of a function, i.e., the "Lagrangian," canonically interpreted as the difference between the kinetic and potential energy at a given time. The balance between the kinetic and potential energies is accounted for by the time integral of the Lagrangian, a scalar function named \textit{action}, which in the following we denote with $\mathscr{A}$. The most striking example of this concise representation is the action of the Standard Model of particle physics (Yang-Mills Theory \cite{Chandra2022}). The equations of motion are then recovered through a cardinal law of physics, the \textit{least action} principle, which is an equivalent to Newton's law of motion. It posits that the path taken by a system between two states is such that $\mathscr{A}$ is stationary. Crucially, $\mathscr{A}$ portrays the system in a defined region of space and for all time periods, and the path can thus arise from any process, both at or out of equilibrium in the statistical mechanics sense. Hitherto, most of the founding models in neuroscience do not yet reconcile with the \textit{least action} principle, and a systematic Lagrangian description of neural dynamics is surprisingly missing. Only indirect evidence has been given\cite{Fagerholm2021,Qiu2014}, e.g., the proof for the Dynamic Causal Modeling (DCM) of Fagerholm et al.\cite{Fagerholm2021}. Reconsidering neural circuits in this light allows neuroscience to formally communicate with seemingly distant fields, borrowing their theoretical schemes and numerical analysis techniques. Given that almost all fields of physics are compatible with this principle, why should neural circuits be an exception? 

\section*{Fundamentals}

\begin{figure}[h!]
\begin{center}
\includegraphics[width=0.7\linewidth]{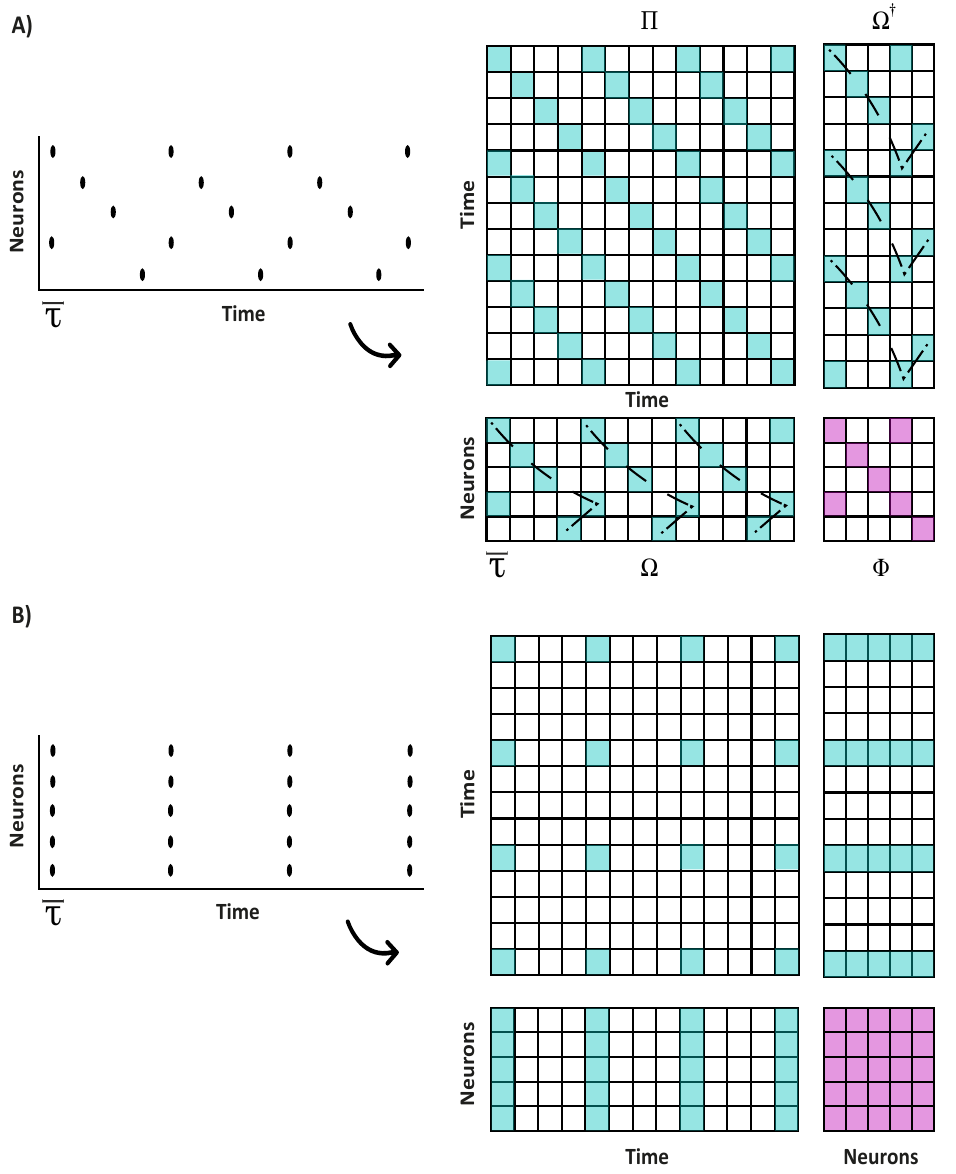}
\caption{A sketch of the spike trains and the hypermatrix fora toy system of 5 neurons recorded for 13 time bins of length $\tau$ for two examples of spatiotemporal patterns. Colors are the same as in the graphical abstract: aquamarine for the kinetic contribution and magenta for the potential one. The elements on the main diagonal of $\Pi$ represent the correlation of the neurons at the same time point (synchronous firing). A peak along the diagonal suggests that the neurons tend to fire together at the same times. Off-diagonal peaks instead indicate consistent lagged relationships between the periodic firings of the ensemble. If the network exhibit specific sequences of periodic firing patterns as in panel A (dashed lines in $\Omega$ and $\Omega{}^{\dagger}$), this will be reflected as periodic structures along both the diagonal and off-diagonal elements of the $\Pi$ matrix. One potential model proposed to explain this and other types of sequential synchronous firing is the synfire chain\cite{Abeles1982,Abeles2004} (see also next section). If groups of neurons fire together periodically in the same time bin (panel B), the $\Pi$ matrix will exhibit multiple distinct periodic peaks that will appear not only on the main diagonal but also off-diagonal. The hypermatrix representation, comprising both the spatial and temporal correlation matrix, provides a compact and complete representation to infer the joint spatiotemporal interactions in neural networks.}
\label{fig:1}
\end{center}
\end{figure}

Our derivation replaces continuous spacetime with a discrete lattice, transforming fields into variables defined on lattice sites at discrete time steps. More formally, we propose a lattice field theory\cite{Buice2007,Buice2013,Wilson1974,Wiese2009,Gupta2011,Zohar2015,Parotto2018, Magnifico2021,Gornitz1992,Peretto2004,Deutsch2004,Singh2020, Franchini2023,Peretto1984,Schneidman2006,Meshulam2024,Qiu2014,BardellaLFT2024} (LFT), the reference computational approach used in QFTs to tackle numerical simulations in regimes where analytical techniques are not feasible. Many authors\cite{Lee1983,Lee1987,Hooft2014,Rovelli1995} have agreed that expressing physical theories in these terms still guarantees the symmetries and conservation laws of continuous formulations. The lattice Lagrangian formalism itself has been studied by multiple authors\cite{DAmbrosio2019,Grimmer2022,Craciun1996,Bourdin2013,Gubbiotti2019,Gubbiotti2020}.
Here, we illustrate the formalism and the basic principles for binary activity, but the arguments can be readily extended to any real-valued signal so as to include the most common neurophysiological measures (Local Field Potentials - LFP -, Multi Unit Acitvity - MUA -, etc..). More formally, this means that our treatment can account for any Potts-like model with multi-spin interactions. Writing the action $\mathscr{A}$ of a neural network using the LFT formalism requires the following simple considerations.\\In computational neuroscience, the accepted view is that the elementary blocks of the neuronal code are electrical impulses called action potentials, or spikes. A neuron emits a spike when it reaches the threshold integration of the electrochemical inputs from other neurons. Hence, by assigning 0 to no spike and 1 to a spike, we can assume that the functional role of a neuron can be described by a binary variable\cite{BardellaLFT2024} whose support is 
\begin{equation}
\Gamma:=\left\{ 0,1\right\}.
\end{equation} Despite its simplicity, this reflects a vast range of complex biochemical interactions in a concise form. We then note that, being intrinsically discrete objects, a set of $N$ neurons can be arbitrarily mapped\cite{BardellaLFT2024} onto the ordered set of lattice vertices 
\begin{equation}
V:=\left\{ 1\leq i\leq N\right\}.
\end{equation}
If we record the activity of the $N$ neurons for a time $T$, we can map the time blocks onto the vertex set 
\begin{equation}
S:=\left\{ 1\leq\alpha\leq T\right\}. 
\end{equation}
The natural temporal ordering crucially fixes the map between time intervals and $\alpha\in S$. The preservation of time ordering is a key feature of our field theoretic approach and is reminiscent of the time-ordered product of field operators used in QFT\cite{LSZ1955}. $T$ can be appropriately discretized according to the minimum time $\tau$ between two computational operations of the neuron. It is convenient to choose the smallest possible value for $\tau$, i.e., the typical duration of a spike: $\tau\approx 1ms$. In this way, since for neural computations, scales below this $\tau$ can be neglected, no temporal information is lost. Neurophysiologists may have noted that $\tau$ is related to the refractory period, which thus yields the minimum relevant time scale, a natural clock time for the system. Indeed, within a $\tau\approx 1ms$, the $i$-th neuron can be reasonably assumed to be either silent (0) or active (1), and its activity at time $\alpha$ expressed with a binary variable $\varphi_{i}^{\alpha}$. The network activity can then be represented with a binary array $\Omega$ of $N$ rows and $T$ columns that we call the \textit{kernel}:\cite{Franchini2023}
\begin{equation}
\Omega:=\left\{ \,\varphi_{i}^{\alpha}\in\Gamma:\alpha\in S,\,i\in V\right\}.
\label{eq:kernel}
\end{equation}
In actual recordings during a neurophysiology experiment, $\Omega$ collects the temporal sequence of spikes of an arbitrary number of neurons aligned with the events or stimuli specific to the chosen experimental paradigm included in $T$. During experiments, we handle multiple trials for the same stimulus presentation or the same behavioral condition. For a single trial, the kernel $\Omega$ simply contains the spike trains for the $N$ neurons recorded. It is established that information encoding and transfer in neural circuits rely on correlations, with brain functions mediated by dynamic changes in the correlated firing of groups of neurons\cite{Gerstein1969,Aertsen1989,Vaadia1995,Shadlen1998,Panzeri1999,Averbeck2006}. Then, let us introduce the space correlation matrix
\begin{equation}
\Phi:=\{\phi_{ij}\in[0,1]:\,i,j\in V\},\ \ \ \ \ \ \ \phi_{ij}:=\frac{1}{T}\sum_{\alpha\in S}\varphi_{i}^{\alpha}\,\varphi_{j}^{\alpha}
\label{eq:phi}
\end{equation}
and the time correlation matrix
\begin{equation}
\Pi:=\{p^{\alpha\beta}\in[0,1]:\alpha,\beta\in S\},\ \ \ p^{\alpha\beta}:=\frac{1}{N}\sum_{i\in V}\varphi_{i}^{\alpha}\varphi_{i}^{\beta}.
\label{eq:pi}
\end{equation}
$\Phi$ contains the pairwise correlations among the $N$ neurons within the whole epoch $T$, while $\Pi$ instead encloses the temporal relationship of the joint occurrences of the ensemble's spikes. These are straightforwardly obtained from the kernel\cite{Franchini2023} through the relations
\begin{equation}
\Omega\,\Omega{}^{\dagger}/T=\Phi,\quad \Omega{}^{\dagger}\Omega/N=\Pi,
\end{equation}
where $\dagger$ indicates the transpose operation. Since the observable matrices can be arranged into a single composite matrix (as shown in the graphical abstract, Figure \ref{fig:1} and Figure \ref{fig:2}), we will call the triplets $\Omega$, $\Pi$, and $\Phi$, the \textit{hypermatrix}.
$\Pi$ gives a time-dependent measure of correlation at different lags, which enables the detection of time structures related to the stimulus presentation, the effect of shared inputs and the associated statistical fluctuations. Each entry represents a delayed interaction of the spike trains of the $N$ neurons during the window $T$, which accounts for the dynamic adjustments of their correlated firing. $\Pi$ is the generalization to $N$ neurons of the traditional Joint Peristimulus Time Histogram matrix\cite{Gerstein1969,Aertsen1989,Vaadia1995} (JPSTH matrix), thus quantifying the cross-correlation as a function of time at the whole network level. The strength of this representation is that, as firstly shown by the seminal works of Gerstein, Aersten and Abeles\cite{Gerstein1969,Aertsen1989,Abeles1982}, the analysis of the profile patterns in $\Pi$ can reveal groups of neurons that form processing units, the so-called cell assemblies. $\Pi$ exposes the sequence of coordinated spiking in which synchronization spreads with a fixed temporal delay from one set of neurons to the next in a temporally identifiable manner. Powerful and fully-automated methods now exist\cite{Schrader2008,Russo2017,Torre2016,Quaglio2018,Tavoni2017,Cocco2011} to reveal this type of correlated activity and trace it back to the responsible cell assemblies. This gives insights on the network information processing and on the possible types of functional connections between the recorded neurons, each of which produces characteristic signatures in the matrix. A sketched example is given in Figure \ref{fig:1}. Finally, it is also essential to model another parameter, the input $I$, which is often unknown or not precisely measurable. For example, $I$ could model the signal arriving from other brain regions to the observed network or the input noise to the same network. In the next section, we shall see how we can properly compute averages to obtain ensemble observables, and how $\mathscr{A}$ can be defined using $\Omega$, $\Pi$, $\Phi$ and $I$.

\section*{Modeling neural networks with the action}

Neural dynamics is expected to follow some evolution influenced by the prior states, i.e., a dynamical process with memory that can be reasonably described by a\\ quantum evolution\cite{Buice2007,Buice2013,Qiu2014,Nanopoulos1995,Friston2010,Crisanti2018,Helias2020,Gosselin2020,Fagerholm2021,Halverson2021,Summers2021,Swan2022,Tiberi2022,Cook2022,Peretto1984,Schneidman2006,Meshulam2024,BardellaLFT2024}.
This may seem peculiar, but it is actually a harmless assumption because the classical (non-quantum) evolution of a system can always be retrieved as a sub-case of the quantum one\cite{Brown2005,LSZ1955,Guerra1975,Parisi1981,Damgaard1987,Parisi1989}. Essentially, we assume that the evolution in time of a system of neurons could be described by a discrete process of interacting binary fields, or \textit{qubits}\cite{Gornitz1992,Peretto2004, Deutsch2004,Singh2020,Franchini2023,Fredkin1982}.
The challenge of determining the temporal evolution of a system of qubits can be remarkably simplified by treating it as a statistical mechanics problem on a lattice\cite{Wilson1974,Wiese2009,Gupta2011, Zohar2015,Parotto2018,Magnifico2021}, which can then be tackled through a spectrum of powerful methods\cite{Wilson1974,Wiese2009,Gupta2011,Zohar2015,Parotto2018,Magnifico2021, DiCastro1969, Balian1974,Wilson1983,Parisi2001,Kadanoff2013,Franchini2021} ({for more references on the LFT and QFT techniques see the section Theoretical insights -TI- or Bardella et al. 2024\cite{BardellaLFT2024}). To model how $\varphi_{i}^{\alpha}$ changes over time, we can use the same language used for elementary particles, a lattice-based statistical mechanics\cite{Wilson1974,Wiese2009,Gupta2011,Zohar2015,Parotto2018, Magnifico2021,Gornitz1992,Peretto2004,Deutsch2004,Singh2020,Franchini2023}. 
Therefore, following the assumptions of QFT\cite{Brown2005,LSZ1955,Guerra1975,Parisi1981, Damgaard1987,Parisi1989,Lee1983,Lee1987,Rovelli1995,Hooft2014,Wilson1974,Wiese2009, Gupta2011, Zohar2015,Parotto2018,Magnifico2021,Gornitz1992,Peretto2004,Deutsch2004,Singh2020, Franchini2023,DiCastro1969,Balian1974,Wilson1983,Parisi2001,Kadanoff2013,Franchini2021}, we postulate the action function\cite{Brown2005,LSZ1955,Guerra1975,Parisi1981,Damgaard1987,Parisi1989} \begin{equation}
\mathscr{A}:\Gamma^{\,\,NT}\rightarrow\mathbb{R}.    
\end{equation}
Denoting the operation of averaging with respect to $\mathcal{A}$ with angle brackets, the ensemble average of the generic observable $\mathscr{O}$ (e.g., the trial-averaged $\Omega$ will be denoted as $\langle\Omega\rangle$) is obtained through a softmax average, which is equivalent to the Gibbs principle applied to $\mathscr{A}$\cite{Guerra1975}, i.e., the principle of \textit{least action}\cite{Brown2005},
\begin{equation}
\langle\mathscr{O}\rangle:=\sum_{\Omega\in\Gamma^{\,NT}}\mathscr{O}\left(\Omega\right)\frac{\exp\left[-\lambda\mathscr{A}\left(\Omega\right)\right]}{\sum_{\Omega'\in\Gamma^{\,NT}}\exp\left[-\lambda\mathscr{A}\left(\Omega'\right)\right]}.
\end{equation}
$\lambda \rightarrow \infty$ yields the classical behavior of the system and is thus identified with the ground state of $\mathscr{A}$\cite{BardellaLFT2024}. In this limiting case, the system becomes conservative, in the sense there are no dissipative dynamics and the path pursued by the system is always the path of least action. This limiting (classical or conservative) case can be understood intuitively in terms of the scaling parameter $\lambda$, which effectively plays the role of a precision or inverse temperature. A simplified action $\mathscr{A}$ can be written by combining $\Omega$, $\Pi$, $\Phi$ and $I$ into a single expression\cite{BardellaLFT2024}:
\begin{equation}
\label{eq:action}
\mathscr{A}(\Omega|\,A,B,I)=\sum_{i\in V}\sum_{\alpha\in S}I_{i}^{\alpha}\varphi_{i}^{\alpha}+T\sum_{i\in V}\sum_{j\in V}A_{ij}\,\phi_{ij}+N\sum_{\alpha\in S}\sum_{\beta\in S}B^{\alpha\beta}p^{\alpha\beta}, 
\end{equation} where $A$ is the matrix of potential interactions and  $B$ is the matrix of kinetic interactions. The concise derivation of Equation \ref{eq:action} can be found in the TI section. The entire information about the system is thus coded in the three observables $\Omega$, $\Phi$ and $\Pi$, the parameters of the theory $A$ and $B$ that control the fluctuations, and the boundary conditions $I$. This simplified action is expected to hold if the synaptic weights can be considered approximately stationary and the observed neurons share the same kinetic properties, e.g. when they approximately share the same firing dynamics if driven by the same input (see TI). We may appreciate the remarkable power of the LFT formulation—its comprehensive nature despite its simplicity. $\mathscr{A}$ succinctly portrays the essentials about the system in a compact form without requiring intricate numerical manipulations. The quantities of the theory are directly mapped into easily accessible and straightforwardly interpretable experimental variables, which is a consistent advantage of this approach.
While frequently employed in neuroscience research to analyze, model and discuss empirical findings, the temporal\cite{Gerstein1969,Aertsen1989,Vaadia1995} and spatial\cite{Panzeri1999,Schneidman2006,Averbeck2006,Okun2015} correlations among neurons have not been previously linked to one another through a physical interpretation. Instead, our LFT approach gives a rigorous yet intuitive understanding of how they contribute to the potential and kinetic energies of a neural system. $\Phi$ contributes to the former and $\Pi$ to the latter, respectively. This provides an easy recipe to relate the spatial correlations to the observed dynamics and vice versa, describing their mutual influence in terms of a fundamental quantity, energy. The \textit{least action} principle then allows to derive all the statistical features needed to determine the properties and functions of a neural network, enabling the formulation of neural interaction theories directly from experimental data (see also Section 3.5 of Bardella et al. 2024\cite{BardellaLFT2024}). In accordance with the triplets $\Omega$, $\Pi$, and $\Phi$, we will call the triplets $A$, $B$, and $I$, the \textit{inverse hypermatrix}.
We can also introduce the covariance matrices, which account for all the information that do not depend on the averages:
\begin{equation}
\label{eq:cov-matrices}
\langle\delta\Pi\rangle:=\langle\Pi\rangle-\frac{\langle\Omega\rangle^{\dagger}\langle\Omega\rangle}{N},\ \ \ \langle\delta\Phi\rangle:=\langle\Phi\rangle-\frac{\langle\Omega\rangle\langle\Omega\rangle^{\dagger}}{T}.
\end{equation}
This converts (non-negative) second order matrices into covariance matrices that quantify coupling in terms of positive and negative covariance by removing the average before computing the correlations in Equations \ref{eq:phi} and \ref{eq:pi}. The parameters $A$ and $B$ can be inferred from these matrices alone\cite{BardellaLFT2024}, which can also be used in combination with $\langle\Omega\rangle$ to reconstruct the input $I$ (see next section).

\section*{Outlooks, implications and limits}

\subsection*{Effective theories for neural recordings}

So far, $N$ includes all the neurons engaged in the neural computation we want to describe, including those that cannot be directly observed. Indeed, in a neurophysiology experiment, we typically have a limited spatial resolution, being able to measure only a fraction of $N$. To relate the microscopic theory of Equation \ref{eq:action} with the marginals on a sparse subset of neurons, we can resort to what in physics is known as an \textit{effective} theory. Effective theories are simplified representations to treat otherwise prohibitive systems\cite{Wells2012} at specific spatial scales or energy ranges while ignoring non-relevant variables. Various areas of physics, from subatomic particles to cosmological structures and network science \cite{Villegas2023}, use effective theories to study how the parameters change when you "zoom in" or "zoom out" on the system, a process known as \textit{renormalization}. Renormalization connects the features at different scales without needing to solve the full theory at all scales. One way of renormalizing is the so-called renormalization by decimation\cite{Wilson1983,Kadanoff2013}, in which the details of the system at small scales are simplified by integrating out most of the degrees of freedom, e.g., spins in magnetic systems, neurons in this context. Then we observe that the brain naturally has spatial symmetries. For example, the brain cortex exhibits highly symmetrical  assemblies of neurons approximately organizing into horizontal layers of columnar groupings\cite{Mountcastle1997,Jones2000,Buxhoeveden2001,Buxhoeveden2002,Cruz2005,Lubke2007}. The arrangement of layers and columns varies in thickness, cell type, and density across different parts of the cortex and for different species. Furthermore, the functional and morphological definitions of columns do not always overlap\cite{Horton2005}, and these structures do not crystallize in time and space, with dynamic changes occurring even on the scale of minutes and hours. Nonetheless, the columnar organization is the most accepted view for the structural and operational components of neural circuitry\cite{Bastos2012}. The so-called minicolumns, on average $\sim 40-50 \mu$m in diameter\cite{Mountcastle1997,Cruz2005}, are deemed to be the basic mesoscale ones. In this paragraph, we refer to minicolumns, meaning vertically oriented and horizontally separated discrete neuronal assemblies\cite{Horton2005,Bastos2012}. Hence, this does not necessarily overlap with the functional definition of a column as a cluster of cells sharing the same tuning properties or receptive field parameters. To show how to exploit such symmetries and apply renormalization, we consider recordings from the Utah array (typically 96 recording channels; Blackrock Microsystems, Salt Lake City), one of the most widely used multi-electrode interfaces. Since thousands of kernels are already available from 20 years of recording with Utah arrays across many species, including human patients\cite{Bougou2024}, we hope to encourage the systematic and joint use of LFTs based on these datasets. Due to its planar geometry, its electrodes that penetrate around 1.5 mm into the cortex, and their pitch ($400 \mu$m), the Utah array is able to record from neurons belonging to horizontally separated minicolumns sampled approximately from the same superficial cortical layer $z$ at a distance sufficient to limit self-interaction terms (see Figure \ref{fig:2}, panel A). Instead, a single shank multi-electrode array with contact points spread out vertically, such as Neuropixels\cite{Steinmetz2021} or SiNAPS\cite{Angotzi2019} probes, would sample across various layers of the same column. 
We name $V_{xyz}$ a 3D volume of tissue containing all the neurons within an average height from the cortex surface $z$ and a $xy$ section in the horizontal plane (Figure \ref{fig:2}, panel A). Here, renormalizing by decimation assumes that if any of the neurons in $V_{xyz}$ emit a spike (are active), the entire lattice cell, and hence the whole minicolumn, is activated. Considering the Utah array specifics, we can organize the $xy$ plane into a sub-lattice $\mathbb{L}'_{2}$ at height $z$ whose step is much greater than the diameter of the individual column, so that the activities recorded at different points belong to adequately spaced minicolumns\cite{BardellaLFT2024}. Using these very general steps, we can model any Utah array recording as a decimated lattice (Figure \ref{fig:2}, panel B). The neural dynamics around each electrode tip are then given by an on/off field $\hat{\varphi}_{x'y'}^{\alpha}$ that identifies the state of the observed minicolumn (see also TI). Therefore, theory is straightforwardly mapped into experimental observations with a decimated kernel
\begin{equation}
\hat{\Omega}:=\{\hat{\varphi}_{x'y'}^{\alpha}\in\Gamma:\,x'y'\in\mathbb{L}'_{2},\,\alpha\in S\}.
\label{eq:experikernel}
\end{equation}
Cortical minicolumns are reduced to a system of $2+\epsilon$ dimensions that could model cortical structures and areas. Note that Equations \ref{eq:phi} and \ref{eq:pi} still hold, and for the decimated kernel $\hat{\Omega}$ they immediately give the experimental hypermatrix (Figure \ref{fig:2} panel D). Therefore, the LFT formalism\cite{Wilson1974,Wiese2009,Gupta2011,Zohar2015, Parotto2018,Magnifico2021,Gornitz1992,Peretto2004,Deutsch2004,Singh2020,Franchini2023} reveals that empirical recordings are comparable to a renormalized field theory, with cortical layers behaving as the interacting fields of elementary particle theory. It is crucial to stress that the effective theory for a generic experimental recording will depend on the chosen renormalization scheme, individual features of the recording interface, the observed network (local circuit and/or area), and the experimental settings. Modeling and computing the correction for effective theories of neural interactions is precisely a component where expertise from nuclear physics could be useful in transferring knowledge to neuroscience. This would not only be a formal artifice, but it would also boost a new generation of models with the entry into neuroscience of established schemes for analyzing, simulating, renormalizing, and, in some cases, exactly solving such theories\cite{DiCastro1969,Balian1974,Wilson1983,Parisi2001,Kadanoff2013,Efrati2014,Franchini2021}.

\begin{figure}[h!]
\begin{center}
\includegraphics[width=0.6\linewidth]{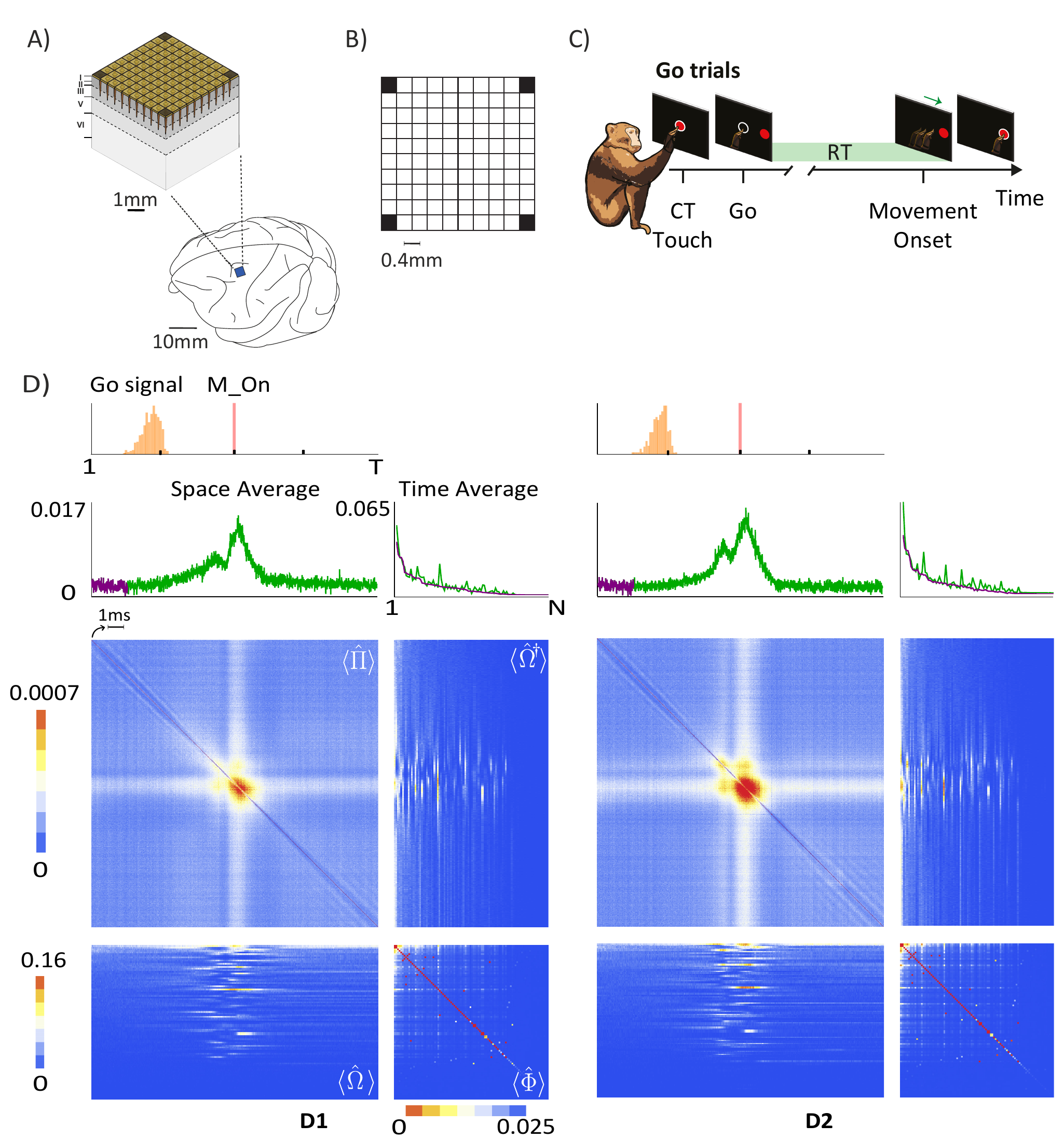}
\caption{Example of application to experimental data. Here \textit{in vivo} recordings from the dorsal premotor cortex (PMd) of non-human primates during a behavioral task (adapted from Bardella et al. 2024 \cite{BardellaLFT2024}). \textbf{A) Cortical minicolumns sampling of the Utah array}: the listening volume of each electrode can be estimated to be approximately the distance between the electrodes ($\sim 400\mu m$)\cite{Hill2014}. In PMd, with around 1.5 mm penetration, the Utah 96 samples activity from around the inner Baillager band\cite{Rapan2021,Opris2011}. \textbf{B) Decimated lattice} of the decimated kernel $\hat{\Omega}$ for Utah 96 interfaces. \textbf{C) Behavioral task} that required visually guided arm movements toward a peripheral target (Go trials) that could appear in two opposite directions (D1 or D2). Monkeys had to reach and hold the peripheral target to get the reward. CT: central target; Go: Go signal appearance; M\_on: Movement onset. \textbf{D) Experimental hypermatrix} averaged over trials for the decimated kernel $\hat{\Omega}$ of Equation \ref{eq:experikernel}. Neural activity is aligned [-1, +1]s ($T = 2s$) around the M\_on to include the distributions of the stimuli (the Go signal, orange distribution and M\_on, magenta). Here the $I$ of Equation \ref{eq:action} represents the time markers for the stimuli presented during the task. Green traces above the $\langle\hat\Pi\rangle$ matrix are the time evolution of the spatially-averaged activity of the network (Space average). Green traces above $\langle\hat\Omega^{\dagger}\rangle$ are instead the time-average activity for each neuron $i$ (Time average). Purple traces are the observables computed in the first 250 ms (“baseline”), which, as expected, are indistinguishable for both conditions. The kernels and $\langle\hat\Phi\rangle$ are sorted according to the activity in the first 250 ms of D1, before the appearance of any Go signal. Black ticks are every 500 ms.}
\label{fig:2}
\end{center}
\end{figure}

\subsection*{Learning neural interactions from data}
Characterizing a neural network's collective response entails measuring average quantities such as mean spiking activity or correlations between neurons. Then, to infer the underlying interactions, one should determine the parameters of a chosen statistical models  based on experimental observations. This is the so-called inverse problem\cite{Swendsen1984} and is a major task (for an excellent survey and references see Nguyen et al. 2017\cite{Nguyen2017}). A convenient approach is to incorporate into the model the minimum number of constraints possible on the network's statistical features. This class of models is known as \textit{maximum entropy models,} and it encompasses various fields of research that have provided multiple strategies to address it and numerous algorithms to tackle model inversion\cite{Nguyen2017}. To give a coarse-grain idea of these techniques, one of the most popular algorithms for finding parameters maximizes the model's average log-likelihood so that the moments of the resulting distribution match a set of specified values\cite{Berger1996}.
A well-known application of maximum entropy models and their related inverse problems to neurobiology is thanks to Schneidman, Tkacik and colleagues \cite{Schneidman2006,Tkacik2014} who inferred spatial couplings of salamander retina neurons assuming an Ising model at equilibrium. Same approach followed by Tavoni, Cocco and colleagues on prefrontal recordings in rats\cite{Tavoni2016,Tavoni2017,Cocco2017} 
However, such a family of models usually considers the distribution of neural activity regardless of its temporal order, constraining only the mean firing rate of the neurons and the pairwise spatial correlations. This means that, typically, only the potential interaction energy contribution is considered. However, a complete description of neural interactions should encompass not only the interaction among circuit's elements but also, and more importantly, its relationship to the observed dynamics. Both are comprised in Equation \ref{eq:action}, which thus generalizes the maximum entropy model in that explicitly contains the time evolution of the system through a kinetic term. Indeed, as also detailed in the TI section, eq \ref{eq:action} decomposes the dependencies of the coupling into space and time, describing not only the influence that neurons located in distinct regions of space have on one another but also how they retain information about their past and how they are coupled over time. In Equation\ref{eq:action} the only constraint on the dynamics is that the state of the system at instant $\alpha$ depends only on the previous $\beta\leq\alpha-1$ so that no retrocausality is allowed. Consequently, beyond the opportunity of directly estimating dynamics from experimental data, the generality of $B$ has major implications for developing generative models based on Equation \ref{eq:action}. For instance, one could simulate neural time series according to a biologically plausible model (e.g., a Leaky Integrate and Fire - LIF -), estimate the distribution of theoretical couplings over simulations according to Equation \ref{eq:action} and compare them with those obtained from experimental data. Likewise, Equation \ref{eq:action} can be used to study the $\mathscr{A}$ of model families with chosen connectivity and dynamics by constraining the matrices $A$ and $B$. This would allow analyzing various known dynamical regimes (e.g., asynchronous, oscillatory, etc..) by classifying parameters as a function of $\mathscr{A}$. This may also be useful to gain knowledge about specific functionalities of neural circuits, for instance, how $\mathscr{A}$ is altered by modifying the excitation/inhibition balance? What is the $\mathscr{A}$ for an "optimal" choice of the parameters? What happens to $\mathscr{A}$ when neuromodulation is altered? In other words, Equation \ref{eq:action} could be used to perform generative modeling consistent with the \textit{least action} principle, accounting for a biologically and physically-inspired description simultaneously. One potential problem with this approach could be parameter degeneracy, i.e, diverse parameter sets produce the same network dynamics. However, strategies are being developed to curb this problem\cite{Lederman2022} and $\mathscr{A}$ itself could be used as a metric to explore the phase space of different candidate models and score families of parameters in simulation studies. Similarly, in experimental studies, having the system's dynamics derived directly from observations and not imposed a priori can be of great advantage for inference methods. For example, some features of the dynamics can be extracted directly from the $\Pi$ matrix, which could be used to restrict the space of the parameter distributions to infer. It would also be possible to use such features to simplify the construction of analytically solvable models to represent realistic circuitry. For instance, to compute the partition function associated to the $\mathscr{A}$ of the synfire chain\cite{Abeles1982,Abeles2004}, the model proposed to explain the sequential spatio-temporal activation at the millisecond-scale of multiple spike patterns. Notably, very recent works\cite{Goncalves2020,Gokmen2021,Fischer2022,Merger2023,Hoover2023,Song2024} have shown how to exploit the power of deep learning to obtain the correspondence between modeling parameters, the statistics of the training data, and the representations artificial networks build of the processed information. The work of Merger et al.\cite{Merger2023} is of particular significance because it makes an explicit connection between the internal representation of a class of generative deep neural network models (i.e., the Invertible Neural Networks - INNs -) and the learned physical theory formulated in terms of $\mathscr{A}$. In their work, an explicit distribution for various types of data and the corresponding microscopic theories of interactions are extracted and interpreted from the parameters of the trained INNs in an unsupervised manner. This, beyond the broad spectrum of applications in neuroscience, also offers promising prospects for improving the interpretability of deep learning models\cite{Ras2022}. In this respect, our derivations significantly alleviate the computational burden in the case of experimentally recorded neural activity, given that they reduce the number of parameters needed to compute $\mathscr{A}$ from $T^{2}N^{2}$ to $N^{2}+T^{2}$ (see TI). From a neuroscience perspective, $A$ in Equation \ref{eq:action} is the connectivity matrix, which can be constrained to have any topology or embody any assumptions about the degree of sparsity. For a general characterization of the dynamics, one can in principle entertain or estimate asymmetric coupling ($A_{ij}\neq A_{ji}$). It is essential to notice that asymmetric coupling of this sort breaks detailed balance and introduces solenoidal (zero divergence) dynamics\cite{Aguilera2022}. This affords the opportunity to model non-equilibrium steady-state solutions, especially in the setting of exogenous input. This becomes a central elements for a proper description of neural activity or, more generally, of biological time series that characteristically show solenoidal dynamics (e.g., oscillations, biorhythms, life cycles and, more generally, stochastic chaos). Our approach is very flexible in that, for instance, it would allow constraining $A$  with adaptable sparsity specifications depending on the scale of analysis. In the case of micro-scale networks, this would comprise, for example, axons adjacencies or labeling of the recorded cell types. At the macroscale, constraints could come from the integration of imaging data or white matter tractography of various cortical areas. This corresponds to incorporating biologically informed connectivity priors\cite{Friston2011} for empirical neural activity, which is becoming common practice for a variety of models in neuroscience\cite{Suarez2020}. A well-known example is DCM which builds large-scale MRI effective connectivity models, scoring the likelihood of various symmetry constraints on the network architecture. Correspondingly, our methods allow us to construct multiscale effective connectivity models with the additional constraint of empirical dynamics\cite{Friston2003,Penny2004,Friston2011,Prando2020}. Another example is integrating anatomical details (i.e., from tract-tracing or gene expression experiments) to build voxel-scale models of the mouse connectome\cite{Knox2019}. Therefore, our formulation enables a multiscale investigation of the interaction between the structural connectivity of a network, its intrinsic dynamics, and the emerging functional relationships among its units. Importantly, $\mathscr{A}$ can be employed within a Bayesian inference framework, similarly to the negative variational free energy in the context of DCM, where it serves as lower bound approximation of the evidence of competing models\cite{Friston2010,Friston2011,Zeidman2019,Zeidmann2023,Friston2003,Penny2004,Prando2020} as well as in the domains of predictive coding\cite{Kawato1999,Rao2002} and active inference applied to neural circuits, where it is used to model how neurons dynamically update their internal states and synaptic weights to predict and infer the causes of their sensory inputs\cite{Friston2006,Parr2017,Palacios2019,Gandolfi2022}.

\subsection*{LFT and biohybrids networks}
Recent progress in the growth, manipulation, and integration of biological brain tissue and bio-inspired electronics devices has opened up intriguing possibilities beyond the conventional use of studying the functioning of natural neural networks in living organisms. Brain organoids\cite{Chiaradia2020,Zheng2022,Sharf2022,Friston2023_organoids} and neuromorphic systems\cite{Obaid2020,Vassanelli2016,Buccelli2019,George2020} are the most striking examples. Currently, there are efficient methods to develop synthetic networks and hybrid circuits that imitate biological systems and connect artificial and real brain networks. These circuits could be programmed using the binary LFT language. Indeed, neuromorphic chips use event-driven processing, meaning that computations are only performed when events (spikes) occur, leading to significant energy savings\cite{Yao2024}. The LFT language could then be used as an efficient paradigm for abstracting computations, performing simulations, and tuning the artificial synaptic weights of the chips based on the activity of the recorded cells during the design and test phase of neuromorphic devices\cite{Serb2020,Keene2020}. In addition, the potential application of transforming organoids and neuromorphic networks into functional neural circuits has been also recently discussed by Zheng and coworkers\cite{Zheng2022}. 
This route seems the most viable practical implementation of utilizing natural (and possibly biohybrid) neurons for conducting physical LFT simulations, eventually realizing the concepts suggested by Halverson\cite{Halverson2021}, but with real neurons.

\subsection*{Limitations}
We conclude discussing some of the open challenges of this new framework. The effective theory, which would actually fit the experimental data, is connected to the microscopic theory (i.e., at the level of single neurons) through extensively studied "renormalization" procedures. However, the precise formulation and calculation of these procedures are still in the early stages and require more research for a comprehensive understanding. The first approximation we employ is the two-body truncation (see TI), which is deemed to be valid under the condition of small covariances. This approximation essentially involves applying a maximum entropy model to $\mathscr{A}$. Hence, it should be, at worst, equivalent to the usual "stationary" max entropy model described above. The second one is implementing additional stationary conditions truncation, which is applicable when the synaptic connections remain constant across the time scale of the empirically recorded brain activity and homogeneous firing dynamics of the neurons in the network. Reasonably, this can be assumed to be valid for the time scales of most neurophysiology experiment. Indeed, a comparison of recording sessions that are considerably distant in time might be a useful way to test this approximation. In the case of single-neuron activity, our approach describes the on-off state of the neurons over time. Therefore, the map between the couplings of Equation \ref{eq:action} and the actual synaptic anatomical structure, or that of the ion channels, is not trivial. However, as discussed above, we argue that it should be possible to reconstruct the parameters of $\mathscr{A}$ from knowledge of the anatomical structures in the LFT formulation. It is still unclear to what extent one can determine the anatomical structures only based on observable dynamics, even if it is reasonable to expect an overlapping similarity for measures obtainable with a sufficient amount of precision, such as the $A$ matrix or the grand covariance (see Equation \ref{eq:grandcov}, TI). Generally speaking, this approach also suffers of the curse of inverse problems, which are, to some extent, ill-posed. This is because an high number of parameters need to be determined from a relatively small subset of observation, with an inverse operator needed to map the measurement vector to the estimated "ground-truth", as for example, in the Magnetoencephalography (MEG) or Electroencephalography (EEG) source localization problem, where statistical models try to assign the true sources activation to the observed measurements.
\newpage

\section*{Theoretical insights}
\label{Theo}

\subsection*{The action of a neural network}

We resume here the fundamentals of the theory. For a comprehensive treatment please refer to Bardella et al. 2024\cite{BardellaLFT2024} and Franchini 2023\cite{Franchini2023}. Following the principles of Statistical Field Theory\cite{Guerra1975,Parisi1981,Damgaard1987,Parisi1989}, we postulate the analytic euclidean action function\cite{Brown2005,LSZ1955,Guerra1975,Parisi1981,Damgaard1987, Parisi1989}$
\mathscr{A}$, that can be written as a Taylor's expansion as follows\cite{BardellaLFT2024}:
\begin{multline}
\mathscr{A}(\Omega|\,F,I):=\sum_{i\in V}\sum_{\alpha\in S}I_{i}^{\alpha}\varphi_{i}^{\alpha}+\sum_{i\in V}\sum_{j\in V}\sum_{\alpha\in S}\sum_{\beta\in S}F_{ij}^{\alpha\beta}\varphi_{i}^{\alpha}\varphi_{j}^{\beta}+\\
+\sum_{i\in V}\sum_{j\in V}\sum_{h\in V}\sum_{\alpha\in S}\sum_{\beta\in S}\sum_{\gamma\in S}F_{ijh}^{\alpha\beta\gamma}\varphi_{i}^{\alpha}\varphi_{j}^{\beta}\varphi_{h}^{\gamma}+
\sum_{i\in V}\sum_{j\in V}\sum_{h\in V}\sum_{k\in V}\sum_{\alpha\in S}\sum_{\beta\in S}\sum_{\gamma\in S}\sum_{\delta\in S}F_{ijhk}^{\alpha\beta\gamma\delta}\varphi_{i}^{\alpha}\varphi_{j}^{\beta}\varphi_{h}^{\gamma}\varphi_{k}^{\delta}+\,...\label{eq:cvcg}
\end{multline}
The terms are the one-, two-, three-, and four--vertex interactions, etc., while the tensors $F$ collects the parameters of the theory. We postulate\cite{Schneidman2006} that terms with more than two vertices can be neglected,
\begin{equation}
F_{ijh}^{\alpha\beta\gamma}=0,\ \ \ F_{ijhk}^{\alpha\beta\gamma\delta}=0,\ \ \ ...
\end{equation}
Therefore, the proposed action reduces to:
\begin{equation}
\mathscr{A}(\Omega|\,F,I)=\sum_{i\in V}\sum_{\alpha\in S}I_{i}^{\alpha}\varphi_{i}^{\alpha}+\sum_{i\in V}\sum_{j\in V}\sum_{\alpha\in S}\sum_{\beta\in S}F_{ij}^{\alpha\beta}\varphi_{i}^{\alpha}\varphi_{j}^{\beta}.
\end{equation}
Notice that is formally equivalent to the max entropy model considered in the works of Schneidman and colleagues\cite{Schneidman2006} but with a key difference: the same neuron at different times is considered as two different neurons. It is the action that ultimately makes them look the same evolving in time. We introduce the grand covariance (see also section 3 of Bardella et al. 2024\cite{BardellaLFT2024}),
\begin{equation}
\label{eq:grandcov}
\mathcal{C}_{ij}^{\,\alpha\beta}:=\langle\varphi_{i}^{\alpha}\varphi_{j}^{\beta}\rangle_{\mu}-\langle\varphi_{i}^{\alpha}\rangle_{\mu}\langle\varphi_{j}^{\beta}\rangle_{\mu}
\end{equation}
from which the couplings can be reconstructed via inference methods\cite{BardellaLFT2024}. We can further simplify the theory by dropping interaction terms where both indices are different, that we improperly call "non-relativistic" approximation because it interprets the columns of the kernel $\Omega$ as sequence of causally ordered isochronic surfaces\cite{BardellaLFT2024}, or Markov blankets\cite{Friston2021}:
\begin{equation}
\mathscr{A}(\Omega|\,A,B,I):=\sum_{i\in V}\sum_{\alpha\in S}I_{i}^{\alpha}\varphi_{i}^{\alpha}+\sum_{i\in V}\sum_{j\in V}\sum_{\alpha\in S}F_{ij}^{\alpha\alpha}\varphi_{i}^{\alpha}\varphi_{j}^{\alpha}+\sum_{\alpha\in S}\sum_{\beta\in S}\sum_{i\in V}F_{ii}^{\alpha\beta}\varphi_{i}^{\alpha}\varphi_{i}^{\beta}
\label{eq:non-rel-action}
\end{equation}
Non relativistic is intended in the "non general" reltivistic sense. Indeed, in Equation 129 of Bardella et al 2024\cite{BardellaLFT2024} we showed how to reduce to the special relativistic Klein-Gordon Lagrangian as a special case.
This approximation is already useful, as it can account for variable synapses and multiple neuron species with "only " $NT(N+T)$ parameters. We show a test of this specific step in Figure 21 of Bardella et al. 2024\cite{BardellaLFT2024}.
Finally, we assume that the terms with $i\neq j$ are stationary in time and those with $\alpha\neq\beta$ are stationary among the considered neurons, i.e. \cite{BardellaLFT2024}
\begin{equation}
F_{ij}^{\alpha\alpha}=A_{ij},\quad F_{ii}^{\alpha\beta}=B^{\alpha\beta},
\end{equation}
which is equivalent to assuming that the connections do not change at the considered time scale and that the observed neurons have all the same kinetic properties. This is reasonable for the scale of most behavioral neurophysiology experiments. Our approximation reduces the number of parameters that should be computed to reconstruct the action from $T^{2}N^{2}$ to $N^{2}+T^{2}$, significantly enhancing the computational tractability.\\ 
The action can be rewritten using the correlation matrices\cite{BardellaLFT2024}:
\begin{equation}
\mathscr{A}(\Omega|\,A,B,I)=\sum_{i\in V}\sum_{\alpha\in S}I_{i}^{\alpha}\varphi_{i}^{\alpha}+T\sum_{i\in V}\sum_{j\in V}A_{ij}\,\phi_{ij}+N\sum_{\alpha\in S}\sum_{\beta\in S}B^{\alpha\beta}p^{\alpha\beta}
\end{equation}
The information is thus coded in the three observables $\langle\Omega\rangle$,
$\langle\Phi\rangle$ and $\langle\Pi\rangle$, which form the hypermatrix\cite{BardellaLFT2024} . Finally, we introduce the covariance matrices of eq. \ref{eq:cov-matrices}. The parameters $A$, $B$ can be inferred from these matrices alone. The covariances can then also be used in combination with $\langle\Omega\rangle$ to reconstruct the input $I$\cite{BardellaLFT2024}. As described in the main text, note that anatomical connections of any topology can be encoded in the interaction matrix $A$, potentially integrating any developments in the mapping of the cortex.

\subsubsection*{Renormalization}
\label{decimated kernel}

To link theory and experimental observations we apply a simple renormalization \cite{DiCastro1969,Wilson1983} scheme based on Franchini 2023\cite{Franchini2023}. For more formal details, please refer to section 3 and 4 of Franchini 2023\cite{Franchini2023} and section 4 of Bardella et al. 2024\cite{BardellaLFT2024}. Following the main text, let us indicate a cubic lattice with $\mathbb{L}_{3}$ and with $z$ ($xyz\in\mathbb{L}_{3}$) the average height from the surface of the cortex at which a given cortical layer is located. Let $xy$ be the position of the center of gravity of the cortical minicolumn section in the horizontal plane. To model the minicolumn layers, we will define a partition of the space $\mathbb{R}^{3}$ into volumes of equal size according to the lattice cells. To simplify, we will approximate the cortical minicolumns as square-based minicolumns. We remark that we use an euclidean reference frame to allow comparisons with existing histological, fMRI, and other structural data\cite{Breitenberg}. However, note that the underlying Euclidean geometry does not, in principle, restrict the LFT parameters.
Also, this may help highlight effects due to possible correlations with Euclidean topology\cite{Buzsaki2012}. The layers of the minicolumns are thus represented by the lattice cells
\begin{equation}
U_{xyz}:=U_{x}U_{y}U_{z}\subset\mathbb{R}^{3}
\end{equation}
Now, calling $v\left(i\right)\in\mathbb{R}^{3}$ the position of the $i-$th neuron (possibly of its cell body), we can group by the volume in which they are located
\begin{equation}
V_{xyz}:=\left\{ i\in V:\,v\left(i\right)\in U_{xyz}\right\}.
\end{equation}
Each of these groups of neurons will have its own associated kernel
\begin{equation}
\Omega_{xyz}:=\left\{ \varphi_{i}^{\alpha}\in\left\{ 0,1\right\} :\,i\in V_{xyz},\,\alpha\in S\right\} .
\end{equation}
Then, one could further group the neurons, first by index $z$, so as to form the cortical minicolumns. The vertices belonging to the minicolumn are
\begin{equation}
V_{xy}:=\bigcup_{z\in\mathbb{L}}\,V_{xyz}
\end{equation}
that is the set of neurons that constitutes the minicolumn at position $xy$. The kernel is
\begin{equation}
\Omega_{xy}:=\{\Omega_{xyz}\in\left\{ 0,1\right\} ^{V_{xyz}}:\,z\in\mathbb{L},\,\alpha\in S\}
\end{equation}
and describes the activity of the single cortical minicolumn in $xy$. It is possible to observe this activity directly through some interfaces, like Neuropixels\cite{Steinmetz2021} or SiNAPS\cite{Angotzi2019} probes or deep multielectrode shanks. The minicolumns are in the end grouped again to form the cortex structures and areas, 
\begin{equation}
V:=\bigcup_{xy\in\mathbb{L}_{2}}\,V_{xy}
\end{equation}
and the original kernel can thus be expressed in terms of the minicolumns:
\begin{equation}
\Omega=\{\Omega_{xy}^{\alpha}\in\left\{ 0,1\right\} ^{V_{xy}}:\,xy\in\mathbb{L}_{2},\,\alpha\in S\},
\end{equation}
so that it represents a two-dimensional lattice of cortical minicolumns, a system in 2+$\epsilon$ dimensions. For the above, we can consider the experimental kernel for a specific tubular layer
\begin{equation}
\Omega_{z}:=\{\Omega_{xyz}^{\alpha}\in\left\{ 0,1\right\} ^{V_{xyz}}:\,xy\in\mathbb{L}_{2},\,\alpha\in S\}.
\end{equation}
The points are organized in a planar sub-lattice $x'y'\in\mathbb{L}'_{2}$ (of the observed cortical layer $z$) whose step is much greater than the diameter of the individual minicolumn, so that the activities recorded at the various points belong with high probability to different and well-spaced minicolumns. 
To model the spacing between the probing points, we apply a renormalization by decimation on $\Omega$, and obtain the decimated activity kernel of eq. \ref{eq:experikernel} of the main text
\begin{equation}
\hat{\Omega}:=\{\hat{\varphi}_{x'y'}^{\alpha}\in\left\{ 0,1\right\} :\,x'y'\in\mathbb{L}'_{2},\,\alpha\in S\},\ \ \ \hat{\varphi}_{x'y'}^{\alpha}:=\mathbb{I}(\Omega_{x'y'z}^{\alpha}\neq0).
\label{eq:Methods_experikernel}
\end{equation} 
Considering potential corrections for systematic errors and approximations, such kernel is intended to model the sensor recording.  According to our arguments, it should be comparable with a renormalized theory. Notice that here renormalization occurs only in space and the information coming from the digitalization of neuronal signals is largely preserved. 

\section*{Resource availability}


\subsection*{Lead contact}


Requests for further information and resources should be directed to and will be fulfilled by the lead contact, Giampiero Bardella (giampiero.bardella@uniroma1.it).

\subsection*{Data and code availability}


Any additional information required to reanalyze the data reported in this paper is available from the lead contact upon request.

\section*{Acknowledgments}


This research was partially supported by Sapienza University of Rome via grant\\ PH11715C823A9528 (to Stefano Ferraina) and RM12117A8AD27DB1 (to Pierpaolo Pani). 
We acknowledge a contribution from the Italian National Recovery and Resilience Plan (NRRP), M4C2, funded by the European Union–NextGenerationEU \\(Project IR0000011, CUP B51E22000150006, “EBRAINS–Italy” (to Stefano Ferraina). \\ We thank Giancarlo La Camera (Stony Brook University), Danilo Benozzo (University of Pavia), Simone Scardapane (Sapienza - University of Rome), Indro Spinelli (Sapienza - University of Rome) and Diego Fasoli (University of Leeds) for constructive comments.

\section*{Author contributions}


Conceptualization, G.B. and S.F. (Simone Franchini); methodology, G.B. and S.F.; writing-–original draft, G.B.; writing-–review \& editing, G.B., S.F., P.P. and S.Fe. (Stefano Ferraina); funding acquisition, P.P. and S.Fe.; resources, P.P. and S.Fe.

\section*{Declaration of interest}


The authors declare no competing interests

\newpage






\bibliography{references}

\bigskip


\end{document}